\documentstyle[]{mn}
\ifnfsstwo

\fi

\ifnfssone
  \newmathalphabet{\mathit}
    \addtoversion{normal}{\mathit}{cmr}{m}{it}
    \addtoversion{bold}{\mathit}{cmr}{bx}{it}

\fi

\ifoldfss

\fi

\loadboldmathitalic
\loadboldgreek
                                                               
\def\msun{\thinspace\hbox{$\hbox{M}_{\odot}$}}

\def\gal{galaxy}
\def\gals{galaxies}

\def\ev{evolution}

\def\for{formation}
\def\sfor{star formation}

\def\cf{cooling flow}

\def\cc{cm$^{-2}$}
\def\ccc{cm$^{-3}$}

\title[Magnetic fields in cooling flows]
       {The effects of  magnetic fields in cold clouds in cooling flows}
\author[A. C. S. Fria\c ca and L.C. Jafelice]
{A. C. S. Fria\c ca$^1$ and L.C. Jafelice$^2$\\
$^1$Instituto Astron\^omico e Geof\'\i sico, USP,
Caixa Postal 9638, 01065-970 S\~ao Paulo, SP, Brazil\\
$^2$Departamento de F\'\i sica Te\'orica e Experimental, UFRN,
Caixa Postal 1641, 59072-970 Natal, RN, Brazil}

\pubyear{1996}

\begin{document}
\maketitle

\begin{abstract}
Large masses of absorbing material are inferred to exist in cooling flows in clusters of galaxies
from the excess X-ray absorption in the spectra of some X-ray clusters.
The absorbing material is probably in the form of  cold clouds pressure--confined
by the surrounding, hot, X--ray emitting gas.
The cold clouds could remain relatively static until they are destroyed by
evaporation or ablation, or give rise to \sfor.
If the final fate of the clouds is stars,
the IMF of the stars formed over the whole \cf\ region ($r\sim 100$ kpc)
should be biased to low masses,
to avoid a very luminous, blue halo for the central galaxy of the \cf.
However,
there is  evidence for bright \sfor\  in the innermost ($r\la 10$ kpc) regions
of some \cf s, 
and, therefore, the biasing of the IMF towards low masses 
should not occur or be less important at smaller radii.
The consideration of magnetic fields may shed light on these two points.
If magnetic fields are present,
the magnetic critical mass should be considered, besides the Jeans mass,
in establishing a natural mass scale for star formation.
When this new mass scale is taken into account, we obtain the
right variation of the biasing of the IMF with the radius
in addition to inhibition of high-mass \sfor\ at large radii.
We also demonstrate that magnetic reconnection is a more efficient mechanism
than ambipolar diffusion
to remove magnetic fields in cold clouds.

\end{abstract}


\section{Introduction}

X-ray observations of  \cf s in clusters of \gals\ have shown evidence
of  large masses of intrinsic X-ray absorbing material
(White et al. 1991; Allen et al. 1993; Allen \& Fabian 1997).
The X-ray spectra show excess photoelectric absorption over that 
detected in our Galaxy, and require an absorbing column of $\sim 10^{21}$
\cc\ covering the core of the cluster out to at least $\sim 100$ kpc.
The absorbing material is probably in the form of cold clouds
embedded in the \cf\ (White et al. 1991; Ferland, Fabian \& Johnstone 1994).
The total amount of cold mass ranges from 
$\sim 10^{11}$ to more than $10^{12}$ \msun.
These masses are in good agreement with those expected to accumulate from
the \cf s if the present deposition rates determined from deprojection
of  X-ray brightness profiles have been maintained 
during several Gyr.
Only smaller masses of gas below X-ray-emitting temperatures have been
derived from observations  at wavelengths other than X-rays.
Up to 10$^8$ \msun\ of ionized gas at $\sim 10^4$ K is present in some
clusters within the inner few kpc, in the form of optical line-emitting
filaments (Heckman et al. 1989), although masses $<10^6$ \msun\ are
more common. 
Molecular gas has been detected in the inner regions of Perseus via
CO emission (Lazareff et al. 1989; Mirabel \& Sanders 1989; Braine et al. 1995).
Also in Perseus, neutral hydrogen has been discovered via absorption
of diffuse radio emission by the 21 cm line (Jaffe 1990).
All of these detections, other than X-ray absorption require cold gas
to be present only at radii $r \la 15$ kpc.
Moreover, apart from the Perseus cluster, only upper limits exist for the
radio observations. 
Atomic hydrogen 21 cm observations limit the mass of optically thin H~I 
to at most $10^9-10^{10}$ \msun\ (McNamara, Bregman \& O'Connell 1990;
Jaffe 1992; O'Dea, Gallimore \& Baum 1995; Dwarakanath, van Gorkom \& Owen 1994).
CO observations yield limits of $10^8-10^{10}$ \msun\
for the mass in molecular hydrogen,
applying the usual Galactic CO luminosity to H$_2$ mass conversion factors
(O'Dea et al. 1994; McNamara \& Jaffe 1994; Braine \& Dupraz 1994;
Antonucci \& Barvainis 1994).
These upper limits are consistent with the $10^{5-6}$ \msun\ of
molecular hydrogen at $T\approx 1000-2000$ K
in the inner $\sim 5$ kpc of central cluster galaxies in \cf s
inferred from recent $K$-band spectroscopy of  central galaxies in \cf s,
which has detected emission lines at $1.8-2.1$ $\mu$m from
H$_2$ (1--0) S(1) through S(5) transitions 
(Jaffe \& Bremer 1997; Falcke et al. 1998).

The final fate of the cooled gas has long been a puzzle, as mass deposition
integrated over a Hubble time typically gives $\sim 10^{12}$ \msun.
The gas could accumulate as cold clouds or collapse to form stars.
It has previously been assumed that the cooled gas forms directly
into low-mass stars with high efficiency
(Fabian, Nulsen \& Canizares 1982; Sarazin \& O'Connell 1983;
Schombert, Barsony \& Hanlon 1993; Kroupa \& Gilmore 1994).
The stars formed could not have
a solar neighbourhood IMF, otherwise the galaxy at the centre of 
the \cf\ would have a very luminous, blue halo.
If the gas ends up as starlike objects, the IMF should be biased
to dim objects as red dwarfs, brown dwarfs or Jupiters.
In fact, the conditions within \cf s are so different from those
in our own interstellar medium that the IMF of the stars formed
is expected to be different from that in the Galaxy.
Widespread cold clouds, however, are a possible sink for the cooled gas, so that
star formation need not be very efficient
(Daines et al. 1994), but the detection of the cold clouds
has eluded both H~I and CO observations
(O'Dea et al. 1994, 1995).

Further problems, therefore, arise
if the gas removed from the hot phase of the \cf\ is in the form of cold clouds.
On the one hand, the clouds must cool very fast to a dense, very cold state, 
otherwise they would
have a number of observational signatures: coronal lines, HI emission
or absorption lines, and CO emission (Voit \& Donahue 1995).
On the other hand, they should be prevented from collapsing very fast
into stars or low-mass objects, otherwise the cold phase would not have
the large masses inferred from observations.
In addition, there is  evidence for massive \sfor\  in the innermost regions
of some \cf s (Allen 1995, Smith et al. 1997), 
which allows one to infer that 
the suppression of  \sfor\ or the biasing of the IMF
towards low masses does not occur or is less important at smaller radii.
Therefore, at least in the inner regions of the \cf, cold clouds are been converted
into stars.

In this paper we investigate the role of magnetic fields on the evolution
of cold clouds in cooling flows, focusing specially on the effects of magnetic
fields on star formation in the cold clouds.
In previous papers (Jafelice \& Fria\c ca 1996, hereafter JF; Fria\c ca et al. 1997), 
we explored the role of magnetic fields as heating source
of the warm phase ($T\sim 10^4$ K) of the \cf\ medium represented by the
optical filaments.
As a matter of fact,
\cf s clusters have important magnetic fields as revealed
by high Faraday rotation measure (RM) observed in some of the most vigorous \cf s
(Ge \& Owen 1993; Taylor, Barton \& Ge 1994).
Gordon et al. (1994) have found that all of the \cf s with high RM at their centres
have optical-line filaments or nuclear emission, which lead they to suggest
that magnetic reconnection (MR) could power the emission lines of the optical filaments.
However, our exploration of both MR (JF)
and Alfv\'en waves (Fria\c ca et al. 1997) as heating/ionizing  source
of the filaments, indicated that these mechanisms
could not meet the energetic requirements
of  the most luminous filament systems.
Nevertheless, magnetic fields can have a number of important consequences 
for the dynamics of  the \cf\ as well for the \ev\ and \for\ of condensations 
arising out of the \cf\ (David \& Bregman 1989; 
Loewenstein  1990; Loewenstein \& Fabian 1990;
Soker, Bregman \& Sarazin 1991; Balbus 1991; Hattori, Yoshida \& Habe 1995;
Christodoulou \& Sarazin 1996; Zoabi, Soker \& Regev 1996, 1998).
Magnetic fields are particularly important in supporting and confining the cold clouds
(Daines et al. 1994).
The shielding due to magnetic fields prevents the clouds from being evaporated
by the surrounding $1-7\times 10^7$ K medium.
On the other hand, magnetic fields should be removed of the cold clouds
in order for magnetic support not to prevent the clouds from rapidly
reaching the very low temperatures needed to elude detection via H~I or CO lines.

Also with respect to the issues of inefficient luminous star formation 
as well as of the variation of the biasing of the IMF with radius,
magnetic fields play an important
role, since the natural mass scale for star formation, the Jeans mass,
should be replaced by the magnetic critical mass
if magnetic fields are present.

In Section 2, we compare the efficiency
of ambipolar diffusion and MR
as mechanisms for removing magnetic fields in cold clouds in \cf s.
Section 3 presents calculations of the evolution of cooling
condensations in \cf s with magnetic pressure and MR
until the formation of cold clouds. 
Section 4 makes predictions on \sfor\ along the \cf\ based on the
magnetic critical mass as a mass scale for star formation.
Our main conclusions are given in Section 5.

\section{Magnetic Support in Cold Clouds}

Magnetic fields in cooling flows represent only a minor
fraction of the ambient thermal pressure,
$\beta = P_B/P_{gas} =(B^2/8\pi)/nk_B T \approx 0.01-0.1$.
(Note that, following the convention used in some studies on cooling flows
[e.g., David \& Bregman 1989; Donahue \& Voit 1991],
we define $\beta = P_B/P_{gas}$, whereas in plasma physics, 
$\beta_{pl}=P_{gas}/P_B$ denotes the ``beta parameter" of the plasma.)
Notwithstanding, they should be taken into account
since magnetic pressure would dominate the dynamics of cold clouds that
condense out of the cooling flow
(David \& Bregman 1989; Zoabi et al. 1998).
As shown in simulations of the evolution of condensations in
cooling flows (JF),
in the beginning of the growing of the perturbation,
the magnetic pressure is unimportant in comparison to the thermal
pressure. 
However, as the gas within the perturbation cools,
under conditions of frozen-in field, the component of the magnetic field
perpendicular to the direction of compression increases ($B \propto \rho$
in plane-parallel geometry), and eventually $\beta$ becomes
$>1$. At this stage, the cloud becomes magnetically supported
and its compression is halted. After a phase of optical emission,
corresponding to the optical filaments, the temperature drops
so much that the optical emission is put out.

These extinguished filaments, whether they harbour star formation or not,
are the end point for the mass removed in cooling flows and could be
identified with the dark, cold clouds invoked by Daines et al. 
(1994) to explain the excess X-ray absorption in cooling flows.
The final fate of the
removed  mass is naturally either faded filaments or
a cooling flow population of mainly low-mass dim objects.
If the strength of the remaining magnetic field in the filaments were high
enough to support
them against further collapse, the cloud could remain relatively warm
due to heating by the surrounding X-ray emitting cooling flow.
However, in this case the gas could be seen either in CO emission
or in the 21 cm line.
In addition, with no cold clouds,
star formation would be suppressed, at variance with observational
evidence in favor of it.
The maintenance of magnetic support also would avoid dust \for.
Dust has been invoked (Hu 1992; Donahue \& Voit 1993) to explain
the line ratios of the optical filaments and
colour maps reveal that dust features are common around
central  \gals\ of \cf s (Sparks, Ford \& Kinney 1993; 
McNamara \& O'Connell 1992, 1993, McNamara et al. 1996a, 
Pinkney et al. 1996).
Dust widespread in the hot intracluster medium has also being suggested
as responsible for the excess soft X-ray absorption in \cf s instead of cold clouds,
in view of the difficulties in eluding infrared, submillimiter, and radio
detections of cold clouds (Voit \& Donahue 1995).
Very cold, dense, dusty clouds have been considered 
(Fabian, Johnstone \& Daines 1994) as candidates for
the absorbers giving the excess X-ray absorption in cooling flows.
Some mechanism is needed to consume the magnetic field almost completely.
In this way, magnetic support is removed, the cloud
collapses and the temperature drops to very low values, thus allowing
the formation of molecules and dust, as well as star formation.

Ambipolar diffusion has been invoked to remove magnetic fields
in cold clouds in cooling flows (Daines et al.1994).
The timescale for this process is (McKee et al. 1993)
\begin{equation}
t_{AD}={3<\sigma v> \over 4\pi G\mu_H m_H} x_i \; , 
\end{equation}
or $t_{AD}=7.3\times 10^{13}x_i$ yr, for the collision rate coefficient
$<\sigma v>=1.5\times10^{-9}$ cm$^3$ s$^{-1}$ and 
$\mu_H=1.4$ ($\mu_H m_H$ is the mass per hydrogen nucleus), 
typical values for a cold molecular gas of solar abundances
(Nakano 1984).
The harder X-rays from the cluster maintain an ionization fraction
$x_i = n_e/n_H \ga 1.5\times 10^{-5}$ in the cloud, even at very low temperatures
(see next section).
The implied timescale $t_{AD} \ga 10^9$ yr is very long.

As a matter of fact, a much more efficient mechanism for removing magnetic field
is MR.
The evolutionary models for optical filaments in \cf s of JF have considered 
MR as the mechanism powering the line emission of the  filaments.
MR is expected to proceed at a rate
\begin{equation}
t_{MR}={l_{MR} \over M_e v_A} \; ,
\end{equation}
where 
$l_{MR}$ is the scale over which the MR takes place
(i.e. the scale over which the magnetic field reverses its direction),
$V_A=B/(4\pi\rho)^{1/2}$ is the Alfv\'en velocity
and $M_e$ is the MR efficiency
($M_e$ is the Alfv\'enic Mach number of the effective velocity $V_e$
of the reconnection).
For $\beta <1$, MR occurs very slowly ($M_e \ll 1$),
however, for  $\beta > 1$, there are several mechanisms of fast MR.
Therefore, fast MR is expected to occur
during the phase of  optical emission of the condensations in \cf s,
when $\beta \gg 1$  due to the rapid cooling of the gas.
For $\beta > 1$,
$M_e \approx 1/\ln R_m$ (Priest 1982), 
where $R_m$ is the magnetic Reynolds number.
From the definition of $R_m$, $R_m=V_e l_{MR}/\eta$
($\eta$ is the magnetic diffusivity )
and from $M_e = 1/\ln R_m$,
$M_e$ is given by the solution of $M_e \ln (M_e V_A l_{MR}/\eta)=1$.
$M_e$ is a very slowly varying function of $V_A l_{MR}/\eta$:
for $V_A l_{MR}/\eta$ increasing from $10^{15}$ to $10^{25}$,
$M_e$ decreases from 0.032 to 0.019.
In conditions typical of the intense optical emission phase ($T \la 10^5$ K)
of the condensations in cooling flows, $M_e \simeq 0.02$ (JF).
Considering
$\beta = 40$,  $v_A=80$ km s$^{-1}$, and $l_{MR}=10$ pc
as representative values of the optical emission phase (JF),
the resulting time scale is $t_{MR}=6\times 10^6$ yr.
This timescale is much shorter than $t_{AD}$ ($\ga 10^9$ yr),
and is comparable both 
to the duration of the purely X-ray emitting phase ($\sim 3\times 10^6$ yr) 
and to that of the optical emitting phase ($> 2\times 10^6$ yr)
of JF models.
Therefore, MR provides an efficient means of
suppressing magnetic support in cold clouds in cooling flows.

\section{Evolution of condensations in the cooling flow
and formation of  cold clouds}

We have investigated the formation of cold clouds
since the formation of condensations out of the hot phase ($T \ge 10^7$ K)
of the cooling flow within the scenario outlined in Section 2.
In order to study optical filaments in \cf s, JF have performed calculations
of cooling condensations in \cf s from $T=10^7$ K to $T=4\times 10^3$ K.
Here, we extend JF calculations down to a temperature of 100 K,
which would allow the formation of cold clouds in the central region
of the condensation.

 \begin{figure}
 \vspace{106mm}
 \caption{The evolution of temperature (solid line), density (dotted line), 
the ratio $\beta$ (dashed line), and the magnetic field (dashed-dotted line) 
in the centre of the condensation
for model A (upper panel) and model B (lower panel).
Note that, following the convention used in some studies on cooling flows,
we define $\beta = P_B/P_{gas}$, whereas in plasma physics, 
$\beta_{pl}=P_{gas}/P_B$ denotes the ``beta parameter" of the plasma.
}
 \end{figure}

The evolution of the cooling condensations is obtained by
solving the hydrodynamical equations of mass, momentum and
energy conservation using a 1D hydrodynamical code
(see Fria\c ca 1993; Fria\c ca \& Terlevich 1998).
Our simulations have been run with plane-parallel geometry
using a Lagragian grid with 300 zones.
The self-gravitation of the condensations is taken into account.
Since there is no ionization equilibrium
for temperatures lower than $10^6$ K,
the ionization state of the gas at $T<10^6$ K is
obtained by solving the time-dependent ionization equations,
for all ionic species of H, He, C, N, O, Ne, Mg, Si, S, Ar and Fe.
We adopt a non-equilibrium cooling function for temperatures
lower than $10^6$ K, since the recombination time of important ions
is longer than the cooling time at these temperatures.
The cooling function and the coefficients of collisional ionization,
recombination and charge exchange of the ionization equations
are all calculated with the atomic database of the photoionization code
AANGABA (Gruenwald \& Viegas 1992).
The adopted abundances are solar (Grevesse \& Anders 1989).
Since the version of AANGABA used in JF did not include molecules,
their cooling function was valid only for $T \ga 5\times 10^3$ K.
In the present extension  of JF calculations to lower temperatures, we use
the equilibrium cooling functions and molecular fractions  down to $T=100$ K,
given by a purely collisional  model (Lepp et al. 1985, and references therein).

The clouds are modelled as slabs shrinking in the direction transverse to 
that of the \cf\ (which is the radial direction towards the cluster centre,
since we assume spherical geometry for the \cf).
The faces of the slab are in the plane of the magnetic field,
so that the magnetic field is perpendicular to the direction of compression.
The initial density perturbations
are characterized by an amplitude $A$ and a length scale $L$,
following $\delta\rho/\rho=A\,sin(2\pi\,x/L)/ (2\pi\,x/L)$,
where $x$ is the direction of the compression (expansion) of the perturbation.
We have also assumed that the perturbations are isobaric and nonlinear ($A$ = 1).
The slab geometry is justified by the fact that the observed line emission is filamentary,
which suggests that the perturbations are sheetlike rather than spherical.
We start to follow the evolution of the perturbations from the nonlinear
stage in view of the uncertainties about processes suppressing
the growth of thermal instabilities in cooling flows.

In this work, we consider two models, representing perturbations evolving
in the inner and in the outer parts of the \cf.
In model A, representing the \ev\ of a cooling condensation
in the inner \cf\ at a radius $r=10$ kpc,
the unperturbed $n_H=0.1$ cm$^{-3}$ and $T=10^7$ K are assumed.
$L$ was fixed at 1 kpc for all the models.
This length scale is suggested, for instance, by the spatial fluctuations in the
velocity of optical filaments resolved in nearby cooling flows (Heckman et al. 1989).
We fixed $\beta=P_B/P_{gas}=0.1$ for the unperturbed medium,
a representative value for the range of $\beta=0.01-1$, expected in the
central 10 kpc of cooling flows.  
We have considered that MR proceeds at an efficiency $M_e=0.02$.
The MR heating  is turned on only for $\beta > \beta_{on}=1$.
The value of $l_{MR}$ used in eq. (2) is derived assuming that
the magnetic field suffered a 1D compression as the gas condensed
from the intracluster medium to form the condensations.
The smoothness of the radio images of radio haloes in clusters imply
a correlation length of the magnetic field $l_c \la 15$ kpc (Tribble 1993).
During a 1D compression, the quantity $l_c n_H$ is conserved.
Assuming $l_{c,ICM}= 10$ kpc and $n_{H,ICM}=10^{-3}$ cm$^{-3}$ 
for the intracluster medium,
$l_{MR}\equiv l_c $ within the condensations is obtained from
the local $n_H$.

Model B, describing  the outer \cf\ ($r=100$ kpc)
has $M_e=0.02$ and  $\beta_{on}=1$, and
unperturbed $L=10$ kpc, $n_H=5\times 10^{-3}$ \ccc, and $T=7\times 10^{7}$ K.
The radial variation of $n_H$ and $T$ assumed in models A and B
follows X-ray spectroscopic studies and image deprojection analysis
which allow to derive temperature gradients and the density runs with radius.
The radial dependence of the density implied by models A and B,
$n_H =5\times 10^{-3} (r/100\; {\rm kpc})^{-1.30}$ \ccc,
 is consistent the averaged radial profile
of $n_H =(4.64 \pm 0.88) \times 10^{-3} (r/100\; {\rm kpc})^{-1.26\pm 0.19}$
\ccc\ found by White, Jones \& Forman (1997) in their sample
of large ($\dot M > 50$ \msun\ yr$^{-1}$) \cf s
detected with the {\sc Einstein Observatory}.
With respect to the temperature gradient,
the cooling flow region $10 \la r \la 100$ kpc separates the inner \cf, where
the gas temperature approaches the virial temperature 
of the central \gal\ (typical $\sigma=300$ km s$^{-1}$ or $T=6.6\times 10^6$ K)
from the general ICM, with a temperature roughly equal 
to the cluster virial temperature
(typical $\sigma=1000$ km s$^{-1}$ or $T=7.5\times 10^7$ K).

For model B, the unperturbed value of $\beta$ was derived from 
the relation of the magnetic field at the inner radius $r_i=10$ kpc ( model A)
to the magnetic field at the outer radius $r_o=100$ kpc (model B)
following Soker \& Sarazin (1990).
We assume: 1) frozen-in field; 2) spherical symmetry for the flow;
and 3) that at the outer radius $r_o$ 
the field is isotropic, i.e. $B_{o,r}^2=B_{o,t}^2/2=B_o^2/3$
and $l_{o,r}=l_{o,t}\equiv l_o$
(where $B_{o,r}$ and $B_{o,t}$ are the radial and transversal
components of the magnetic field $B_o$
and $l_{o,r}$ and $l_{o,t}$ are the coherence length of the
large-scale field in the radial and transverse directions).
In the discussion of Soker \& Sarazin (1990), the inward cooling
flow is assumed to be homogeneous. 
However, detailed studies of the observed X-ray brightness profile of \cf s
have revealed that $\dot M$ is not constant with radius,
but that it decreases towards the centre 
following approximately $\dot M \propto r$
(Thomas, Fabian \& Nulsen 1987).
Therefore, we modified the calculation of the magnetic field 
of Soker \& Sarazin (1990) by considering 
an inhomogeneous \cf\ (i.e. $\dot M_i \not= \dot M_o$).
Since 
we are modelling  the condensations responsible for removing the mass
in the flow as slabs containing the magnetic field, 
and since 
the magnetic field lines become increasingly radial as the gas flows inward,
we assume that the condensations are parallel to the radial direction.
For a homogeneous inflow and frozen-in field,
$l_{r,i}=l_{r,o}(u_i/u_o)$ ($u$ is the inflow velocity),
and $l_{t,i}=l_{t,o}(r_i/r_o)$,
but in inhomogeneous flow, in which magnetic field is being removed
from the flow by condensations compressed in the transverse direction,
the $l_{t,i}-l_{t,o}$ relation has to be modified to
$l_{t,i}=l_{t,o}(r_i/r_o)( \dot M_i / \dot M_o )^{-1/2}$.
The two components of the field are then given by
\begin{equation}
B_{i,r}^2={1 \over 3} B_o^2 \biggl( {r_i \over r_o} \biggr)^{-4}
\biggl( {\dot M_i \over \dot M_o} \biggr)^2
\end{equation}
and
\begin{equation}
B_{i,t}^2={2 \over 3} B_o^2 \biggl( {r_i \over r_o} \biggr)^{-2}
\biggl( {\dot M_i \over \dot M_o}\biggr) \biggl( {u_i \over u_o} \biggr)^{-2}\; .
\end{equation}
Assuming that $\dot M \propto r$ (and from $\dot M=4\pi r^2 \mu_H m_H n_H u$),
the field strength $B_o$ is given by
\begin{equation}
B_o^2=3 B_i^2 \biggl[ \biggl( {r_i \over r_o} \biggr)^{-2}
+2 \biggl( {r_i \over r_o} \biggr) \biggl( {n_{H,i} \over n_{H,o}} \biggr)^2 \biggr]^{-1}\; .
\end{equation}
The unperturbed values of $n_H$,  $T$, and $\beta$ of model A 
($n_{H,i}=0.1$ \ccc,  $T_i=10^7$ K, and $\beta_i=0.1$)
imply $B_i=28.2$ $\mu$G,
and, from eq. (5) and unperturbed $n_H$ and  $T$ of model B
($n_{H,o}=5\times 10^{-3}$ \ccc\ and  $T_o=7\times 10^7$ K), 
we obtain $B_o=3.64$ $\mu$G and $\beta=4.76\times 10^{-3}$
for model B.

The hydrogen column density of hydrogen nuclei $N_H$  is 
$4.90\times 10^{20}$ \cc\ for model A and 
$2.45\times 10^{20}$ \cc\ for model B.
Note that
for a planar cloud, under hydrostatic equilibrium, the pressure $P_c$
in the midplane is given by 
$P_c-P_0=(\pi/2)G\Sigma^2$,
where $\Sigma$ is the surface mass density,
and $P_0$ is the external pressure, or
$(P_c-P_0)/k_B=4.3\times 10^3 N_{21}$ \ccc\ K,
where $N_H=10^{21}N_{21}$ \cc, implying 
$(P_c-P_0)/k_B=2.1\times 10^3(1.1\times 10^3)$ \ccc\ K for model A(B).
Since $P_0/k_B=2.3\times 10^6(8\times 10^5)$ \ccc\ K for model A(B),
the cooling condensations are confined by the pressure in the surrounding
\cf, not by their self-gravity.

Other implication of our column densities is that the surrounding X-ray flux
from the \cf\ maintains a significant degree of ionization at the centre of the
cloud, even when very low temperatures are attained.
The model of Ferland et al. (1994), describing a slab embedded in the
radiation field typical of a \cf\ with inflow rate of 100 \msun\ yr$^{-1}$
at the cooling radius of 100 kpc, predicts 
an ionization fraction $x_e \simeq 1.5\times 10^{-5}$
at a depth of $N_H=1-2.5\times 10^{20}$ \cc.
Since in both our models A and B, the surrounding \cf\ environment 
is more vigorous ($nT=2.3\times 10^6$ and $nT=8\times 10^5$, respectively) 
than in Ferland et al. model ($nT\approx 3\times 10^5$), 
the highest X-ray flux would lead to even higher ionization levels.

Figure 1 shows
the evolution of temperature, density, the ratio $\beta$, 
and magnetic field strength
in the innermost cell, for models A and  B,
since the beginning of the optical emitting phase (defined when 
the temperature in the innermost cell drops below $5\times 10^5$ K).
The duration of the first, purely X-ray emitting phase
(when $T>5\times 10^5$ K throughout the condensation)
shows little dependence
on the efficiency of MR: it varies from $3.2\times 10^6$ ($1.57\times 10^9$)
yr to $3.3\times 10^6$ ($1.58\times 10^9$) for model A (B)
as $M_e$ varies from 0 to 0.02.
Here we  focus our discussion on the late (beyond the purely X-ray emitting phase)
\ev\ of the condensation,
and, in general, the times will be counted from the beginning of the optical phase.
The duration of the optical phase 
(during which the filament most strongly emits optical lines)
can be estimated from the time for the gas in the centre of the condensation
(the values of all quantities discussed in this section 
are given in the central region of the condensation)
to cool from $T=5\times 10^5$ K to $T<5\times 10^3$ K:
$1.4\times 10^7$ yr and $2.4\times 10^7$ yr for models A and B, respectively.
Note that, although the duration of optical emitting phase 
is similar for both models,
the purely X-ray emitting phase lasts much longer for model  B
than for model  A, for which the two timescales are comparable.
Due to this fact,
the condensations in the inner \cf\ are much more
efficient emitters of optical lines than the condensations in the outer \cf,
which spend a negligible span of their lifetime in the optical phase.

For model A,
the centre of the condensation takes $1.66\times 10^7$ yr to cool from
$5\times 10^5$ K to 100 K.
One can distinguish three stages in the late evolution of the condensation:
a hot stage (with $2\times 10^4 < T < 5\times 10^5$ K);
a warm stage ($4000 < T < 2\times 10^4$ K);
and a cold stage ($100 < T < 4000$ K).
These stages correspond to the presence or not of thermal instability:
the hot stage is thermally unstable, the warm stage corresponds
to the thermally stable regime around $T\sim 10^4$ K,
and the cold stage include the unstable region 
between $\sim 2000$ and $\sim 4000$ K, in which the cooling function
falls rapidly with temperature.
As a consequence, the filament shows
a very short ($10^5$ yr) hot stage,
a long ($1.41\times 10^7$ yr) warm stage,
and a relatively short ($2.5\times 10^6$ yr) cold stage.
The most slowly varying quantity is the magnetic field, which
first rises from an initial value 105 $\mu$G to 136 $\mu$G at $t=8.5\times 10^5$ yr,
and then decreases to a final value 88.8 $\mu$G.
The \ev\ of the magnetic field shows fluctuations of a factor $\sim 1.5$ around
the average $B$.
The fluctuation of a factor $2-2.5$ at $t=1.4\times 10^7$ yr is due to the gas
having reached the thermally unstable part ($2000 \la T \la 4000$ K)
of the cooling function.
$\beta$ shows a peak value of  180 at $t=2\times 10^5$ yr, and then
decreases to a local minimum of 13.3 at $t=1.4\times 10^7$ yr.
Shortly after this time, 
the gas reaches the $2000 \la T \la 4000$ K unstable domain,
and $\beta$ suddenly rises  to $\simeq 55$ due to the rapid drop
of the temperature and the increase of density to keep pressure equilibrium.
At the end of the cold stage, $\beta$ returns to the $\sim 10$ level.
It is important to note that in model A, MR was unable to consume
the magnetic field, (i.e. to reduce $\beta$ to $\la 1$), at least down to a temperature
of 100 K. The nearly constancy of the magnetic field (only a
15 \% decrease from $5\times 10^5$ K to 100 K) suggests that the value of $B$
at even lower temperatures is not significantly lower than the value at $T=100$ K,
and, therefore, that magnetic pressure keeps dominating over thermal pressure.

For model B, the core of the condensation cools from
$5\times 10^5$ K to 100 K in $2.46\times 10^7$ yr.
The duration of  the hot, warm, and cold stages are
$3\times 10^5$ yr, $2.4\times 10^7$ yr and $6\times 10^5$ yr, respectively.
The MR heating is more effective in model B,
and the stable, warm stage begins closer to $T=3\times 10^4$
than to $ T=2\times 10^4$ as in model A 
(in model B, $T$ decreases from $3\times 10^4$ K to $2\times 10^4$ K
in $4\times 10^6$ yr).
The values of $B$ and $\beta$ are lower than in model A.
Again, the more slowly varying quantity is $B$, which a decrease
of  35\% from $B=58.9$ at $t=0$ to 39.9 $\mu$G
at the end of the calculations.
A maximum $B=64$ $\mu$G is reached at $t=6.4\times 10^6$ yr.
$\beta$ has a maximum of 115 at $t=10^6$ yr, and then decreases until
it becomes $\approx 1$ at $t\simeq 2.4\times 10^7$ yr.
At this point,
the temperature plummets from $T\simeq 7000$ K to $T\simeq 300$ K
because the magnetic pressure support has been removed, and the density
has increased drastically to maintain the pressure equilibrium,
thus reducing the cooling time.
Moreover, the gas has reached the thermally unstable region 
of the maximum of the cooling function between $\sim 2000$ K and $\sim 4000$ K.
The spike in $\beta$ and the smaller one in the magnetic field
are the result of the sudden compression of the gas, leading to a temporary
overpressure in the magnetic field, which is soon consumed by MR.
The high density reached by the gas explains the short duration
of the cold stage.
In contrast with model A, 
in model B the magnetic support is removed by MR when $T\sim 7000$ K,
and, after the temporary increase in $\beta$ in the unstable region
$2000 \la T \la 4000$ K, a situation with $\beta \approx 1$ is established.

A comparison between the time scales for cooling ($t_c$) and for MR ($t_{MR}$)
allows one to understand the reason of MR having consumed magnetic energy
in model B down to $\beta \sim 1$ while $\beta$ remains $\sim 10$
in model A. 
The two time scales set the pace for the decrease of temperature and
$\beta$, respectively.
As shown in Section 4, the ratio $t_{MR}/ t_c $ is smaller
in model B than in model A since the onset of MR,
during the purely X-ray emitting phase of the \ev\ of the perturbation.
As a consequence, during this early evolutionary phase,
the magnetic pressure was already more rapidly consumed
in model B than in model A (the initial values of  $\beta$ in the optical
phase are 151 and 100 for model A than for model B, respectively). 
During the optical phase, the trend of  more efficient magnetic field
consumption in model B than in model A persists,
as one can see from the smaller values of $t_{MR}/t_c$ for model B:
at the beginning of the warm stage 
(then $T=2.9\times 10^4$ K for model A, and $T=1.9\times 10^4$ K for model B),
$t_{MR}/t_c=99 (60)$ for model A (B),
and when the temperature drops to $T=10^4$ K
(at $t=8\times 10^6$ yr and $1.4\times 10^7$ yr, for models A and B, respectively),
$t_{MR}/t_c=55 (34)$ for model A (B).
In addition, during the optical phase, model A becomes closer
to the ionization equilibrium than model B, as we can see from
the comparison of the $n_{H\,I}/n_H$ ratio at $T=10^4$ K
of model A (96.8\%) and model B (44.5\%) to the equilibrium value (99.7\%).
As a consequence, the cooling function of model A approaches the
equilibrium cooling function, which is higher than the extreme non-equilibrium
cooling function (i.e. that obtained in conditions of pure cooling without heating),
thus reducing the cooling time.
This shows the importance of a proper calculation of the cooling function
from the actual ionization state of the gas:
the non-equilibrium plasma is over-ionized for its kinetic temperature
because the recombination time of important ions 
was comparable to the evolutionary time of the perturbation,
and in this case, the actual cooling rate is lower than the
equilibrium value because the highly excited species cannot be
efficiently excited by impact.

It should be noted that
for the high values of $\beta$ typical of the inner \cf, MHD instabilities can occur,
which can have interesting effects on the \ev\  of  the perturbation
that have not been considered in this paper.
In particular,
if the cloud is part of another structure, depending on the curvature radius
of the structure in which the perturbation is embedded, 
magnetic tension can play an important role.
Then, in the evolution and motion of the perturbation, besides
other factors --- radiative cooling, thermal and magnetic pressure, buoyancy, drag ---
magnetic tension should also be considered (Zoabi et al. 1996).
For instance, if the perturbation is located in the lower part of a U-shaped
magnetic flux tube (formed from the reconnection of two neighbouring radially elongated
flux loops), the magnetic tension force could uplift the filament from the
inner region of the \cf\ to a few $\times$ 10 kpc from the cluster center
(Zoabi et al. 1996).
Another possible situation is that of a perturbation  located in the lower part of
a loop of magnetic flux inflowing with the \cf\  (Zoabi et al. 1998).
Then, the denser, lower part of the loop cools till catastrophic cooling,
when MR occurs, freeing the lower segment, which falls inward and
may become an optical filament, and eventually ends up as cold clouds.

\section{Star Formation in Magnetized Cold Clouds}

\begin{table*}
\begin{minipage}{160mm}
\caption{Cooling flow and cold cloud parameters}
\begin{tabular}{@{}lcccccccccc}
Model & $r$ & $n_H$ & $T$ & $B$ & $n_{H,c}$ & $B_c$ 
& $\beta_c$ & $M_J$ & $M_B$ & $t_{MR}/t_c$ \\
 & (kpc) & (\ccc) & ($10^7$ K) & ($\mu$G) & ($10^3$ \ccc) & ($\mu$G)
 & & (\msun) & (\msun) & \\[3pt]
A & 10 & 0.1 & 1 & 28.3 & 3.02 & 88.8 & 13.2 & 0.24 & 1243 & 4.29 \\
B & 100 & 0.005 & 7 & 3.64 & 7.94 & 39.9 & 0.96 & 0.40 & 16.3 & 1.84 \\
\end{tabular}
\end{minipage}
\end{table*}

There is substantial evidence in favor of star formation in cooling flows.
The blue light excess over that expected from the underlying galaxy
that has been observed in many massive cooling flows (Johnstone et al. 1987;
Romanishin 1987; McNamara \& O'Connell 1989; Allen et al. 1992;
Crawford \& Fabian 1993; Crawford et al. 1995)
has been interpreted as due to young,
massive stars (Johnstone et al. 1987; McNamara \& O'Connell 1989).
Although an alternative explanation of the blue light excess in terms of
scattered nucleus emission (McNamara \& O'Connell 1993; Crawford \& Fabian
1993) is also consistent with the data, 
at least in the case of a few clusters,
$e.g.$ Perseus (Rubin et al. 1977; Shields \& Filippenko 1990),
Abell 2199 (Bertola et al. 1986), 
and Hydra A (Hansen, Jorgensen \& Norgaard-Nielsen 1995,
Melnick, Gopal-Krishna, Terlevich 1997),
this emission is clearly stellar since strong Balmer absorption lines
can been seen.
Star \for\ is also suggested by  Mg$_2$ index and by  4000-\AA\ break measurements,
which indicate recent \sfor\ in central \gals\ of  \cf s with optical filaments
(Cardiel, Gorgas \& Arag\'on-Salamanca 1995, 1998).

In addition,  simple modelling of the optical spectra of
the central galaxies of clusters with \cf s,
indicate the presence of $\ga 10^6$ O stars in the central $\sim 10-20$ kpc
of the most massive systems (Allen 1995).
In some massive \cf s, the central galaxy contain large numbers of Wolf-Rayet stars,
which provides evidence for the formation of very massive stars ($M \ga 30$ \msun).
The U-B color excess in the centre of the \cf\ allow to infer star formation
rates ranging from a few to a few tens of  \msun\ yr$^{-1}$
in the central $\sim 10$ kpc of the \cf (McNamara \& O'Connell 1989, 1993).
Although some massive star formation is occurring
in the inner regions of the \cf,
there are a number of stringent observational limits
on massive star formation throughout the \cf\ (see Fabian 1994), so that
a star formation with a standard IMF (Scalo 1986) is excluded.
The star formation
should be very inefficient or skewed toward low mass stars
(or even brown dwarfs and Jupiters)
in the $r \ga 10$ kpc region,
otherwise the central galaxy would be too luminous and blue.

The physical conditions in cooling flows are very different from those
in the Galaxy and, as a consequence, it is expected that the IMF will
differ from that of the solar neighbourhood. In particular, the pressure
in the cooling flow is $\sim 100$ times larger than that of our interstellar
medium. As a result, the Jeans mass (McKee et al. 1993)
\begin{equation}
M_J=11.5{(T/10\;{\rm K})^2 \over (nT/10^3 \;{\rm cm}^{-3}{\rm K})^{1/2}}
\; \msun
\end{equation}
is reduced by a factor of $\sim 10$ with respect to the typical value
for the solar neighbourhood. In this way, the IMF is skewed towards
small masses (Fabian, Nulsen \& Canizares 1982; Sarazin \& O'Connell 1983). 
This then explains why so little star formation is ever
seen around central cluster galaxies: the stars that are formed are
too faint to be easily detected.

The above reasoning, however, does not take into account magnetic fields,
which, as shown in Section 2, dominate the dynamics of the cold clouds.
The primary effect of magnetic fields in cold clouds is
to stabilize them against gravitational collapse (Loewenstein \& Fabian 1990).
In addition,  the comparison of the observed IMF with
calculations of the IMF including several instability criteria,
shows that magnetic fields are needed to reproduce
both the position of the peak and the shape of the IMF
(Ferrini, Palla \& Penco 1990).
In the presence of magnetic fields, the Jeans mass must be replaced by
the critical mass $M_B$, taking into account effects of magnetic pressure,
as the mass scale regulating the process of star formation.
For a spherical cloud, $M_B$ can be written (McKee et al. 1993) as
\begin{equation}
M_B=16.2 {(B/10 \; \mu{\rm G})^3 \over (n_H/ 10^3 {\rm cm}^{-3})^2}
\; \msun . 
\end{equation}
For a frozen-in field, magnetically subcritical clouds (those with $M<M_B$)
can never undergo gravitational collapse, whereas magnetically
supercritical clouds ($M>M_B$) can never be prevented from collapse by
the magnetic field.
This distinction has led Shu, Lizano, \& Adams (1987) to suggest that
magnetically subcritical clouds are associated with low mass star formation,
which is observed to occur at relatively low efficiency,
whereas magnetically supercritical clouds form stars with
a relatively high efficiency, and may preferentially form high mass stars.
The later mechanism would explain the fact that
the value of $M_B$ in the Galaxy, $M_B\sim500$ \msun, 
inferred both for the diffuse ISM as for clumps in giant molecular clouds
(McKee et al. 1993),
is close to the upper cutoff $M_{up}$ of the Galactic IMF, 
$M_{up} \approx 100$ \msun. 
In this way, $M_B$ would set the mass scale of the most massive stars,
being tipically a few times $M_{up}$.

In order to investigate the star formation in cooling flows,
we consider the relevant quantities in the \cf\ environment
--- hydrogen density $n_H$,  temperature $T$, magnetic field strength $B$ ---
at two positions in the cooling flow,
$r=100$ kpc (outer cooling flow),  and $r=10$ kpc (inner cooling flow). 
Table 1 shows the  values of $n_H$,
$T$ and $B$ at these radii, for the models A and B in Section 3.
We also show the density $n_{H,c}$, the magnetic field $B_c$,
and the ratio $\beta_c$
at the centre of the condensation when the central temperature drops to 100 K.
Then, the values of $M_J$ and $M_B$ are calculated to assess the
nature of the star formation.
$M_J$ is calculated in the usual simplistic way, by assuming pressure equilibrium
with the surrounding \cf, i.e., by setting $nT$ in eq. (6) as the pressure in the \cf.
$M_B$ is properly calculated,
by using in eq. (7) $B_c$ and $n_{H,c}$ in the cold core of the perturbation.

Daines et al. (1994) claim that the large masses of cold gas found in cooling
flows via X-ray absorption could be explained in terms of inefficient
star formation in the outer ($r\sim 100$ kpc) regions of cooling flows, whereas
a relatively efficient star formation is still allowed in the inner
regions ($r\sim 10$ kpc), where optical filaments are observed
and there is evidence of a cooling flow star population.
Previous argumentation of the reduction of star formation rate
in cooling flows relied on the calculation of the Jeans mass as the
mass scale regulating the typical mass of the stars formed.
However, this reasoning does not account for the differences between
star formation in the inner and the outer cooling flow, since
the difference in pressure is only a factor of  $\sim 3$, implying a small
variation in the Jeans mass ($M_J \propto P^{-1/2}$) and in the sense
opposite of the expected one, that is, higher Jeans masses in the
outer cooling flow (see Table 1).

A solution to this dilemma is given by the fact that, in the presence of
magnetic fields, $M_B$, the critical mass for the collapse of a
magnetized cloud, is the parameter regulating the star formation
at least in the high mass end of the IMF.
In the case of cooling flows, $M_B$ in the inner regions,
$M_B=1243$ \msun, is comparable to the Galactic value of $\sim500$ \msun\
and, therefore, the star formation is expected to be not so different from
that in the Galaxy, that is, the IMF will include massive stars.
On the other hand, in the outer regions of the cooling flow 
$M_B=16.3 $ \msun, which is much smaller than the value in the Galaxy
and, therefore, the star formation must have characteristics very different
from the Galactic one, implying an inhibition of massive star formation or a
displacement of the masses of the objects formed to low values, typical
of red dwarfs, brown dwarfs or Jupiters. 
The considerations above can be the missing link in the scenario in which
there are differences of star formation efficiency between the outer
and the inner cooling flow.

Table 1 also shows the ratio between $t_{MR}$ and the cooling time
$t_c=(5/2) \, kT/n\Lambda(T)$
when the initial perturbation has cooled from the 
environment \cf\ temperature enough to reach $\beta=1$ 
(then, model A shows
$n_H=0.23$ \ccc, $T=2.34\times 10^6$ K, and $B=66$ $\mu$G,
and model B
$n_H=5.1\times 10^{-2}$ \ccc, $T=3.42\times 10^6$ K, and $B=37$ $\mu$G).
At this point, MR is turned on for the first time.
As we see from Table 1,
the MR is more efficient in model B than in model A.
For this reason, in model B, MR has consumed very efficiently 
the magnetic field, and $\beta \approx 1$ is reached before
the temperature has dropped to 100 K.

\section{Conclusions}

We have calculated typical timescales of MR and ambipolar 
diffusion in conditions prevailing in 
cold clouds formed in \cf s and conclude that MR
is far more important than ambipolar diffusion in removing magnetic fields
($t_{MR} \ll t_{AD}$ in these regions).

Since magnetic fields are present in the cold clouds, 
the magnetic critical mass $M_B$ should be considered, besides the Jeans mass,
as a mass scale for \sfor.
We found that in outer part the \cf\ region ($r\sim 100$ kpc),
$M_B$ is low ($\sim 10$ \msun), so predicting an inefficient formation of
massive stars over that region, in agreement with the absence of young star
signature over the \cf\ region as a whole.
In addition, in the innermost regions ($r \sim 10$ kpc) of the \cf,
$M_B$ reaches values close to the Galactic one, implying that the IMF
is plausibly not very different from that in the solar neighbourhood,
and, therefore, that massive stars are formed, as it is indicated by observations 
of young stars (type A or earlier) in the central regions of \cf s.
In this way, the variation of the biasing of the IMF 
with radius to the centre of the \cf\ in the right sense
is reproduced by considering $M_B$ as the mass scale regulating the \sfor.

It should be mentioned that another process which can cause the \sfor\ being
concentrated in the inner regions of \cf s is the interaction of radio jets with
the ICM.
In fact, imaging of central galaxies of \cf s, 
including observations in the $U$-band
(the most effective for studying blue stellar populations)
reveals a variety of blue structure around the central galaxy (McNamara 1997).
Four morphological types can be defined: unresolved point; disk;  lobe;
amorphous. The case for an association between radio sources and \sfor\ is
stronger for the lobe class, characterized by blue lobes of optical continuum 
located several kpc from the \gal\ nucleus. The archetypes of this class, A1795
and A2597, have radio sources with bright lobes of blue continuum along the edges of
their radio lobes (McNamara \& O'Connell 1993; McNamara et al. 1996a,b).
Several characteristics of  the blues lobes support their origin in \sfor: 
1) the low blue polarization (McNamara et al. 1996b); 
2) the absence of detailed correspondence between the blue lobes and the radio lobes;
3) HST images showing what appears to be blue star clusters 
along the edges of the radio lobes (McNamara et al. 1996a).

In the radio triggered \sfor\ model, the radio jets collect 
pre-existent cold clouds
along the edge lobes creating a gas overdensity which triggers or
enhances \sfor, which is located in
the few central kpc of the \cf, where the jets are confined to.
By contrast,
in our model, the \sfor\ occurs during the process of  formation
of cold clouds from perturbations arising from the hot phase of the \cf,
but massive \sfor\ is allowed only in the central region of the \cf.
Within the classification scheme of  McNamara (1997)
for blue structures in cooling flows, the amorphous type,
which encompasses most of the objects and which shows no
obvious association with radio sources, could be explained
by the model presented in this paper.

\section*{Acknowledgments}

We thank an anonymous referee for comments which greatly improved
the presentation of the results of this paper.
A.C.S.F. acknowledges partial support from the Brazilian agency CNPq.
A.C.S.F. and L.C.J. acknowledge partial support from the Brazilian Ministry
of Science and Technology through the program PRONEX/FINEP.

\bsp

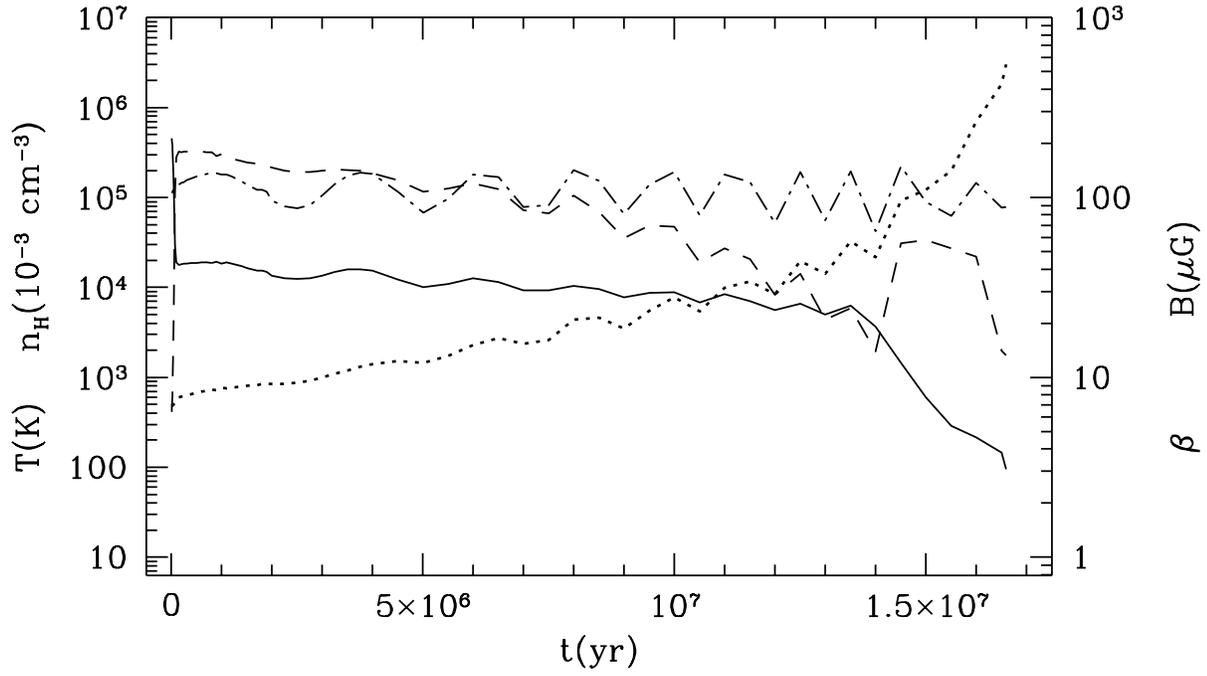
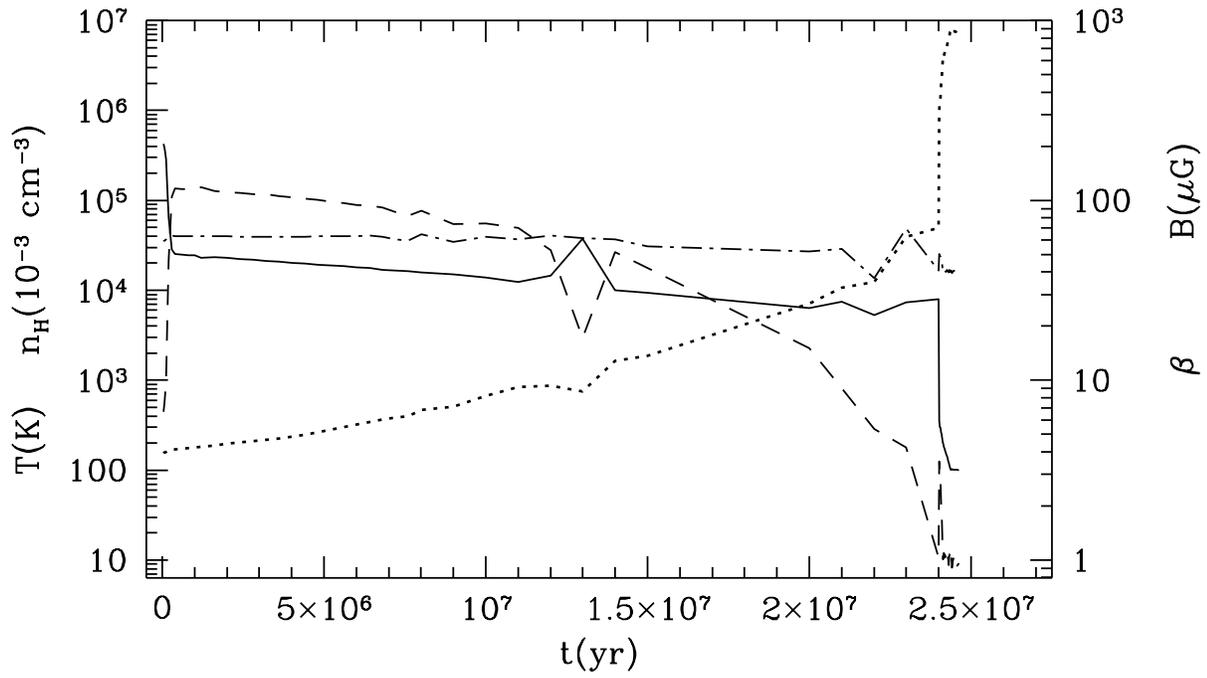
\end{document}